\documentclass[preprint,amsmath,amssymb,aps,showkeys,showpacs]{revtex4}
\usepackage[english]{babel}
\usepackage{graphicx}
\usepackage{graphics}
\usepackage{amsmath}
\usepackage{dcolumn}
\usepackage{amssymb}
\usepackage{bm}
\usepackage[latin1]{inputenc}

\begin{document}
\title{Bound states for a Coulomb-type potential induced by the interaction between a moving electric quadrupole moment and a magnetic field}
\author{K. Bakke}
\email{kbakke@fisica.ufpb.br}
\affiliation{Departamento de F\'isica, Universidade Federal da Para\'iba, Caixa Postal 5008, 58051-970, Jo\~ao Pessoa, PB, Brazil.}

\begin{abstract}
We discuss the arising of bound states solutions of the Schr\"odinger equation due to the presence of a Coulomb-type potential induced by the interaction between a moving electric quadrupole moment and a magnetic field. Furthermore, we study the influence of the Coulomb-type potential on the harmonic oscillator by showing a quantum effect characterized by the dependence of the angular frequency on the quantum numbers of the system, whose meaning is that not all values of the angular frequency are allowed.
\end{abstract}


\maketitle

\section{Introduction}

The interaction between electric and magnetic fields and multipole moments has attracted a great deal of studies, such as the arising of geometric quantum phases \cite{pesk,eab,hmw,zei,zei2,anan2,spa2,whw,chen,b7,bent}, the holonomic quantum computation \cite{ac2,ac3,bf26} and the Landau quantization \cite{lin,b,bf25}. In particular, the electric quadrupole moment has been investigated in systems of strongly magnetized Rydberg atoms \cite{prlquad}, since these systems are characterized by a large permanent electric quadrupole moment. The electric quadrupole moment has also been investigated in molecular systems \cite{quad1,quad} and atomic systems \cite{nucquad,nucquad2}. Besides, recent studies of the interaction between a moving electric quadrupole moment and external fields \cite{chen,b7} have shown a difference between the field configuration that yields the arising of geometric phases for an electric charge \cite{pesk,eab,ab}, an electric dipole moment \cite{hmw,whw} and a moving electric quadrupole moment. In short, this difference arises from the structure of the gauge symmetry involving each system, for instance, the Aharonov-Bohm effect \cite{ab} and the electric Aharonov-Bohm effect \cite{pesk,eab} are based on the gauge symmetry called $\mathcal{U}\left(1\right)$ gauge group, the He-McKellar-Wilkens effect \cite{hmw} and the scalar Aharonov-Bohm effect for a neutral particle with a permanent electric dipole moment \cite{bf26} are based on the $\mathcal{SU}\left(2\right)$ gauge symmetry, while the electric quadrupole moment system can be described by $\mathcal{U}\left(1\right)$ gauge symmetry \cite{bf25}. Moreover, the field configuration that gives rise to the appearance of geometric phases for a moving electric quadrupole moment depends on the structure of the electric quadrupole tensor \cite{chen,b7}. This dependence of the structure of the electric quadrupole tensor has also been pointed out in Refs. \cite{bf25,b7} by dealing with the Landau quantization and the confinement analogous to a quantum dot. 

In this work, we explore this dependence of the field configuration that interacts with the electric quadrupole moment on the structure of the electric quadrupole tensor. We consider a moving electric quadrupole moment and study the arising of bound states solutions of the Schr\"odinger equation due to the presence of a Coulomb-type potential induced by the interaction between a moving electric quadrupole moment and a magnetic field. Furthermore, we study the influence of the Coulomb-type potential on the harmonic oscillator by showing a quantum effect characterized by the dependence of the harmonic oscillator frequency on the quantum numbers of the system, which means that not all values of the angular frequency are allowed.

The structure of this paper is: in section II, we start by making a brief review of the quantum dynamics of a moving electric quadrupole moment interacting with external fields. In the following, we obtain the energy levels corresponding to having a quantum particle subject to an attractive Coulomb-type potential induced by the interaction between a moving electric quadrupole moment and a magnetic field; in section III, we study the influence of the Coulomb-type potential on the harmonic oscillator; in section IV, we present our conclusions.

\section{Coulomb-type potential induced by the interaction between a moving electric quadrupole moment and a magnetic field}

In this section, we discuss the arising of bound states solutions of the Schr\"odinger equation that describes the quantum dynamics of a moving electric quadrupole moment interacting with a magnetic field. We show that the interaction between a moving electric quadrupole moment and a magnetic field can give rise to a Coulomb-type potential, where both scattering and bound state solutions of the Schr\"odinger equation can be obtained. We start with the description of the quantum dynamics of a moving electric quadrupole moment interacting with magnetic and electric fields as proposed in Ref. \cite{chen}. By following Ref. \cite{chen}, we consider an electric quadrupole moment as a scalar particle, then, the potential energy of a multipole expansion in the classical dynamics of an electric quadrupole moment (in the rest frame of the particle) is given by $U=q\,\Phi-\vec{d}\cdot\vec{\nabla}\Phi+\sum_{i,j}Q_{ij}\,\partial_{i}\,\partial_{j}\Phi\cdots$, where $q$ is the electric charge, $\vec{d}$ is the electric dipole moment, $Q_{ij}$ is the electric quadrupole moment (the tensor $Q_{ij}$ is a symmetric and traceless tensor) and $\Phi$ is the electric potential. By considering $q=0$, $\vec{d}=0$ and $\vec{E}=-\vec{\nabla}\Phi$ ($\vec{E}$ is the electric field), the potential energy can be rewritten as $U=-\sum_{i,j}Q_{ij}\,\partial_{i}\,E_{j}$.

If we consider a moving particle (the electric quadrupole moment), we have that the particle interacts with a different electric field $\vec{E}'$. Thereby, the Lagrangian function of this system in the frame of the moving particle is given by $\mathcal{L}=\frac{1}{2}mv^2-\sum_{ij}Q_{ij}\,\partial_{i}\,E_{j}'$. By applying the Lorentz transformation of the electromagnetic field, we have that the electric field $\vec{E}'$ must be replaced by $\vec{E}'=\vec{E}+\frac{1}{c}\vec{v}\times\vec{B}$ up to $O\left(\frac{v^{2}}{c^{2}}\right)$. Now, we have that the field $\vec{E}$ and $\vec{B}$ are the electric and magnetic fields in the laboratory frame, respectively. In this way, the Lagrangian function becomes
\begin{eqnarray}
\mathcal{L}=\frac{1}{2}m\,v^2+\vec{Q}\cdot\vec{E}-\frac{1}{c}\,\vec{v}\cdot\left(\vec{Q}\times\vec{B}\right),
\label{1.2}
\end{eqnarray}
where we defined the components of the vector $\vec{Q}$ by $Q_{i}=\sum_{j}Q_{ij}\,\partial_{j}$ ($Q_{ij}$ is a symmetric and traceless tensor) as done in Ref. \cite{chen}. With the canonical momentum being $\vec{p}=m\,\vec{v}-\frac{1}{c}\left(\vec{Q}\times\vec{B}\right)$, the classical Hamiltonian of this system becomes
\begin{eqnarray}
\mathcal{H}=\frac{1}{2m}\left[\vec{p}+\frac{1}{c}(\vec{Q}\times\vec{B})\right]^2-\vec{Q}\cdot\vec{E}.
\label{1.3}
\end{eqnarray}

In order to proceed with the quantization of the Hamiltonian, we replace the canonical momentum $\vec{p}$ by the Hermitian operator $\hat{p}=-i\hbar\vec{\nabla}$. In this way, the quantum dynamics of a moving electric quadrupole moment can be described by the Schr\"odinger equation \cite{chen,b7}
\begin{eqnarray}
i\hbar\frac{\partial\psi}{\partial t}=\frac{1}{2m}\left[\hat{p}+\frac{1}{c}(\vec{Q}\times\vec{B})\right]^2\,\psi-\vec{Q}\cdot\vec{E}\,\psi.
\label{1.4}
\end{eqnarray}

From now on, we work with the units $\hbar=c=1$. Thereby, let us consider the non-null components of the tensor $Q_{ij}$ being 
\begin{eqnarray}
Q_{\rho z}=Q_{z\rho}=Q,
\label{2.0}
\end{eqnarray}
where $Q$ is a constant $\left(Q>0\right)$. Let us consider the presence of a magnetic field proposed in Ref. \cite{lin}
\begin{eqnarray}
\vec{B}=\frac{\lambda_{m}\rho}{2}\,\hat{\rho},
\label{2.1}
\end{eqnarray} 
where $\lambda_{m}$ is a magnetic charge density, $\rho=\sqrt{x^{2}+y^{2}}$ and $\hat{\rho}$ is an unit vector in the radial direction. The field configuration given in Eq. (\ref{2.1}) was proposed in Ref. \cite{lin} in order to study the possibility of achieving the Landau quantization for neutral particles with a permanent electric dipole moment. This field configuration proposed in Ref. \cite{lin} is based on the He-McKellar-Wilkens effect \cite{hmw}, where the wave function of a neutral particle with a permanent electric dipole moment acquires a geometric phase when the neutral particle encircles a line of magnetic monopoles. In recent years, this radial magnetic field has been achieved through noninertial effects \cite{b}. In this work, we show that the field configuration (\ref{2.1}) cannot yield the Landau quantization for a moving magnetic quadrupole moment, but can give rise to bound states solutions analogous to having a quantum particle confined in the Coulomb potential.

From the interaction between the field configuration given in Eq. (\ref{2.1}) and the magnetic quadrupole moment (\ref{2.0}), then, the Schr\"odinger equation in cylindrical coordinates becomes
\begin{eqnarray}
i\frac{\partial\psi}{\partial t}=-\frac{1}{2m}\left[\frac{\partial^{2}}{\partial\rho^{2}}+\frac{1}{\rho}\frac{\partial}{\partial\rho}+\frac{1}{\rho^{2}}\,\frac{\partial^{2}}{\partial\varphi^{2}}+\frac{\partial^{2}}{\partial z^{2}}\right]\psi-i\,\frac{Q\lambda_{m}}{2m\rho}\,\frac{\partial\psi}{\partial\varphi}+\frac{Q^{2}\lambda_{m}^{2}}{8m}\,\psi.
\label{2.5}
\end{eqnarray}

Note that the operators $\hat{p}_{z}=-i\partial_{z}$ and $\hat{L}_{z}=-i\partial_{\varphi}$ commute with the Hamiltonian of the right-hand side of (\ref{2.5}), then, a particular solution of Eq. (\ref{2.5}) can be written in terms of the eigenvalues of the operator $\hat{p}_{z}$, and $\hat{L}_{z}$: 
\begin{eqnarray}
\psi\left(t,\rho,\varphi,z\right)=e^{-i\mathcal{E}t}\,e^{i\,l\,\varphi}\,e^{ikz}\,R\left(\rho\right),
\label{2.6}
\end{eqnarray}
where $l=0,\pm1,\pm2,\ldots$, $k$ is a constant, and $R\left(\rho\right)$ is a function of the radial coordinate. Thereby, substituting the solution (\ref{2.6}) into Eq. (\ref{2.5}), we obtain the following radial equation:
\begin{eqnarray}
R''+\frac{1}{\rho}R'-\frac{l^{2}}{\rho^{2}}R-\frac{\delta}{\rho}\,R+\zeta^{2}\,R=0,
\label{2.6a}
\end{eqnarray}
where we have defined the following parameters
\begin{eqnarray}
\zeta^{2}&=&2m\mathcal{E}-k^{2}-\frac{Q^{2}\lambda_{m}^{2}}{4};\nonumber\\
[-2mm]\label{2.7}\\[-2mm]
\delta&=&Q\,\lambda_{m}\,l.\nonumber
\end{eqnarray}

 Now, let us discuss the asymptotic behavior of the radial equation (\ref{2.6a}). For $\rho\rightarrow\infty$, we have 
\begin{eqnarray}
R''+\zeta^{2}\,R=0.
\label{2.8}
\end{eqnarray}
Therefore, we can find either scattering states $\left(R\cong e^{i\zeta\rho}\right)$ or bound states $\left(R\cong e^{-\tau\rho}\right)$ if we consider $\zeta^{2}=-\tau^{2}$ \cite{mello,bb2,bb4}. Note that the fourth term on the left-hand side of Eq. (\ref{2.6a}) plays the role of a Coulomb-like potential. This term stems from the interaction between the magnetic field (\ref{2.1}) and the electric quadrupole moment defined in Eq. (\ref{2.0}). Our intention is to obtain bound state solutions, then, the term proportional to $\delta$ behaves like an attractive potential by considering the negative values of $\delta$, that is, $\delta=-\left|\delta\right|$ \cite{bb2,mello,bb4}. This occurs by considering either $\lambda_{m}>0$ and $l<0$ and $\lambda_{m}<0$ and $l>0$ (we consider $Q$ being always a positive number). Observe that these condition forbids the quantum number $l$ to have the value $l=0$, that is, for $l=0$ there are no bound states solutions. Thereby, we rewrite Eq. (\ref{2.6a}) in the form:
\begin{eqnarray}
R''+\frac{1}{\rho}\,R'-\frac{l^{2}}{\rho^{2}}\,R+\frac{\left|\delta\right|}{\rho}\,R-\tau^{2}\,R=0.
\label{2.9}
\end{eqnarray}

Next, by making a change of variables given by $r=2\tau\rho$, we have in Eq. (\ref{2.9}):
\begin{eqnarray}
R''+\frac{1}{r}\,R'-\frac{l^{2}}{r^{2}}\,R+\frac{\left|\delta\right|}{2\,\tau\,r}\,R-\frac{1}{4}\,R=0.
\label{2.10}
\end{eqnarray}
We can  obtain a regular solution for the second-order differential equation (\ref{2.10}) at the origin by writing
\begin{eqnarray}
R\left(r\right)=e^{-\frac{r}{2}}\,r^{\left|l\right|}\,F\left(r\right).
\label{2.11}
\end{eqnarray}
Substituting (\ref{2.11}) into (\ref{2.10}), we obtain the following second-order differential equation
\begin{eqnarray}
r\,F''+\left[2\left|l\right|+1-r\right]F'+\left[\frac{\left|\delta\right|}{2\tau}-\left|l\right|-\frac{1}{2}\right]F=0,
\label{2.12}
\end{eqnarray}
which corresponds to the Kummer equation or the confluent hypergeometric equation \cite{abra}. In this way, the function $F\left(r\right)$ corresponds to the Kummer function of first kind which is defined as
\begin{eqnarray}
F\left(r\right)=\,_{1}F_{1}\left(\left|l\right|+\frac{1}{2}-\frac{\left|\delta\right|}{2\tau},\,2\left|l\right|+1,\,r\right).
\label{2.13}
\end{eqnarray} 

A finite radial wave function can be obtained by imposing the condition where the confluent hypergeometric series becomes a polynomial of degree $n$ (where $n=0,1,2,\ldots$). This occurs when $\left|l\right|+\frac{1}{2}-\frac{\left|\delta\right|}{2\tau}=-n$ \cite{abra,landau}. In this way, from Eq. (\ref{2.7}) we can take $\left(-\tau^{2}\right)=2m\mathcal{E}-k^{2}-\frac{Q^{2}\lambda_{m}^{2}}{4}$ and the parameter $\delta$, thus, we obtain
\begin{eqnarray}
\mathcal{E}_{n,\,l}=-\frac{1}{8m}\frac{\left(Q\,\lambda_{m}\,l\right)^{2}}{\left[n+\left|l\right|+1/2\right]^{2}}+\frac{k^{2}}{2m}+\frac{Q^{2}\lambda_{m}^{2}}{8m}.
\label{2.15}
\end{eqnarray}

Equation (\ref{2.15}) correspond to the energy levels for bound states yielded by a Coulomb-like potential induced by the interaction between the electric quadrupole moment given in Eq. (\ref{2.0}) and the magnetic field defined in Eq. (\ref{2.1}). Observe that the energy levels (\ref{2.15}) are defined for $l\neq0$ as we have discussed above. For $l=0$, there are no bound states solutions because the parameter $\delta$ defined in Eq. (\ref{2.7}), which gives rise to a term that plays the role of a Coulomb-like potential, vanishes. Besides, the energy levels of the bound states (\ref{2.15}) could not be achieved if we have considered the non-null components of the tensor $Q_{ij}$ being, for instance, $Q_{\rho\varphi}=Q_{\varphi\rho}\neq0$ and the field given in Eq. (\ref{2.1}). In this case, the term $\left(\vec{Q}\times\vec{B}\right)\cdot\vec{p}$ present in Eq. (\ref{1.4}) does not induce a Coulomb-type term, thus, no bound states solutions analogous to having a quantum particle confined to a Coulomb potential can be achieved. Hence, the arising of the Coulomb-type potential in Eq. (\ref{2.5}) from the interaction between the magnetic field (\ref{2.1}) and the electric quadrupole moment depends on the structure of the tensor $Q_{ij}$. An analogous case has been pointed out in Ref. \cite{bf25} for the Landau quantization for a moving electric quadrupole moment.  

Further, it should be interesting to explore the quantum dynamics of a moving electric quadrupole moment interacting with external field in the presence of linear topological defects \cite{moraes}. As shown in Ref. \cite{moraes} that both curvature and torsion can modify electric and magnetic fields, in turn, one should expect that the spectrum of energy (\ref{2.15}) can also be modified by the presence of defects.

\section{Influence of the Coulomb-like potential on the Harmonic Oscillator}

In the following, we discuss the influence of the Coulomb-like potential induced the interaction between the electric quadrupole moment given in Eq. (\ref{2.0}) and the magnetic field given in Eq. (\ref{2.1}) on the two-dimensional harmonic oscillator $V\left(\rho\right)=\frac{1}{2}m\omega\rho^{2}$. By following the steps from Eq. (\ref{2.5}) to Eq. (\ref{2.7}), we have the following radial equation
\begin{eqnarray}
R''+\frac{1}{\rho}\,R'-\frac{l^{2}}{\rho^{2}}\,R-\frac{\delta}{\rho}\,R-m^{2}\omega^{2}\,\rho^{2}\,R+\zeta^{2}\,R=0,
\label{3.1}
\end{eqnarray}

Now, we do not need to consider $\zeta^{2}=-\tau^{2}$ as in the previous section, thus, we can consider a general case involving the parameter $\zeta^{2}$ defined in Eq. (\ref{2.7}). Again, the presence of the Coulomb-type term in Eq. (\ref{3.1}) imposes that the angular quantum number is defined for $l\neq0$. Thereby, let us consider a new change of variables given by: $\xi=\sqrt{m\omega}\,\rho$. Thus, we have
\begin{eqnarray}
R''+\frac{1}{\xi}\,R'-\frac{l^{2}}{\xi^{2}}\,R+\frac{\alpha}{\xi}\,R-\xi^{2}\,R+\frac{\zeta^{2}}{m\omega}\,R=0,
\label{3.2}
\end{eqnarray}
where we have defined the following parameter
\begin{eqnarray}
\alpha=\frac{Q\,\lambda_{m}\,l}{\sqrt{m\omega}}=\frac{\delta}{\sqrt{m\omega}}.
\label{3.2a}
\end{eqnarray}

Hence, a regular radial wave function at the origin can be obtained by writing the solution of the second order differential equation (\ref{3.2}) in the form:
\begin{eqnarray}
R\left(\xi\right)=e^{-\frac{\xi^{2}}{2}}\,\xi^{\left|l\right|}\,H\left(\xi\right).
\label{3.3}
\end{eqnarray}
Substituting (\ref{3.3}) into (\ref{3.2}), we obtain
\begin{eqnarray}
H''+\left[\frac{2\left|l\right|+1}{\xi}-2\xi\right]H'+\left[g+\frac{\alpha}{\xi}\right]H=0,
\label{3.4}
\end{eqnarray}
where $g=\frac{\zeta^{2}}{m\omega}-2-2\left|l\right|$. The function $H\left(\xi\right)$, which is solution of the second order differential equation (\ref{3.4}), is known as the Heun biconfluent function \cite{heun,eug,bm,bb2,bb4}:
\begin{eqnarray}
H\left(\xi\right)=H\left[2\left|l\right|,\,0,\,\frac{\zeta^{2}}{m\omega},\,2\alpha,\,\xi\right].
\label{3.5}
\end{eqnarray} 

In order to proceed with our discussion about bound states solutions, let us use the Frobenius method \cite{arf,f1}. Thereby, the solution of Eq. (\ref{3.5}) can be written as a power series expansion around the origin:
\begin{eqnarray}
H\left(\xi\right)=\sum_{j=0}^{\infty}\,a_{j}\,\xi^{j}.
\label{3.11}
\end{eqnarray} 

Substituting the series (\ref{3.11}) into (\ref{3.5}), we obtain the recurrence relation:
\begin{eqnarray}
a_{j+2}=\frac{\alpha}{\left(j+2\right)\,\left(j+1+\theta\right)}\,a_{j+1}-\frac{\left(g-2j\right)}{\left(j+2\right)\,\left(j+1+\theta\right)}\,a_{j},
\label{3.12}
\end{eqnarray}
where $\theta=2\left|l\right|+1$. By starting with $a_{0}=1$ and using the relation (\ref{3.12}), we can calculate other coefficients of the power series expansion (\ref{3.11}). For instance,
\begin{eqnarray}
a_{1}&=&\frac{\alpha}{\theta}=\frac{Q\,\lambda_{m}\,l}{\sqrt{m\omega}\left(2\left|l\right|+1\right)};\nonumber\\
a_{2}&=&\frac{\alpha^{2}}{2\theta\left(1+\theta\right)}-\frac{g}{2\left(1+\theta\right)}\label{3.13}\\
&=&\frac{\left(Q\,\lambda_{m}\,l\right)^{2}}{2m\omega\left(2\left|l\right|+1\right)\left(2\left|l\right|+2\right)}-\frac{g}{2\left(2\left|l\right|+2\right)}.\nonumber
\end{eqnarray}

Bound state solutions correspond to finite solutions, therefore, we can obtain bound state solution by imposing that the power series expansion (\ref{3.11}) or the Heun Biconfluent series becomes a polynomial of degree $n$. Through the expression (\ref{3.12}), we can see that the power series expansion (\ref{3.11}) becomes a polynomial of degree $n$ if we impose the conditions \cite{f1,bb2,bb4,bm,eug}:
\begin{eqnarray}
g=2n\,\,\,\,\,\,\mathrm{and}\,\,\,\,\,\,a_{n+1}=0,
\label{3.13a}
\end{eqnarray}
where $n=1,2,3,\ldots$, and $g=\frac{\zeta^{2}}{m\omega}-2\left|l\right|-2$. From the condition $g=2n$, we can obtain the expression for the energy levels for bound states:
\begin{eqnarray}
\mathcal{E}_{n,\,l}=\omega\left[n+\left|l\right|+1\right]+\frac{Q^{2}\lambda_{m}^{2}}{8m}+\frac{k^{2}}{2m}.
\label{3.14}
\end{eqnarray}

Equation (\ref{3.14}) is the energy levels of the two-dimensional harmonic oscillator under the influence of the Coulomb-type potential induced by the interaction between the electric quadrupole moment given in Eq. (\ref{2.0}) and the magnetic field given in Eq. (\ref{2.1}). Note that the influence of the Coulomb-like potential makes that the ground state to be defined by the quantum number $n=1$ instead of the quantum number $n=0$ as it is well-known. Moreover, we have that the spectrum of energy (\ref{3.14}) is defined for all values of the quantum number $l$ that differ from zero, that is, for $l\neq0$. In the case $l=0$, there is no influence of the Coulomb-like potential induced by the interaction between the electric quadrupole moment and the magnetic field because the parameter $\alpha$ vanishes (as we can see in Eqs. (\ref{3.2a}) and (\ref{2.7})). However, at first glance, the energy levels (\ref{3.14}) do not depend on the parameter $\delta$ defined in Eq. (\ref{2.7}), which gives rise to a Coulomb-type potential. In this way, we would not have the complete information of the spectrum of energy. However, observe that we have not analysed the condition $a_{n+1}=0$ given in Eq. (\ref{3.13a}) yet. We have that this condition yields an expression involving specific values of the angular frequency and the parameter $\delta$, that is, it yields an expression involving the angular frequency and the quantum numbers $\left\{n,\,l\right\}$ \cite{bb2,bm,eug}. A similar analysis has been done in Ref. \cite{bb2}, where a relation involving the harmonic oscillator frequency, the Lorentz symmetry-breaking parameters and the total angular momentum quantum number is obtained. In Ref. \cite{bm}, a relation involving a coupling constant of a Coulomb-like potential, the cyclotron frequency and the total angular momentum quantum number in semiconductors threaded by a dislocation density is obtained. Moreover, a relation involving the mass of a relativistic particle, a scalar potential coupling constant and the total angular momentum quantum number is achieved in Ref. \cite{eug}.

Thereby, in order to obtain a relation between the harmonic oscillator frequency and the Coulomb-type potential induced by the interaction between the electric quadrupole moment and a radial magnetic field, let us assume that the angular frequency $\omega$ can be adjusted in such a way that the condition $a_{n+1}=0$ is satisfied. This means that not all values of the angular frequency $\omega$ are allowed, but some specific values of $\omega$ which depend on the quantum numbers $\left\{n,\,l\right\}$, thus, we label $\omega=\omega_{n,\,l}$. In this way, the conditions established in Eq. (\ref{3.13a}) are satisfied and a polynomial solution for the function $H\left(\zeta\right)$ given in Eq. (\ref{3.11}) is obtained \cite{eug}. As an example, let us consider $n=1$, which corresponds to the ground state, and analyse the condition $a_{n+1}=0$. For $n=1$, we have $a_{2}=0$. The condition $a_{2}=0$, thus, yields 
\begin{eqnarray}
\omega_{1,\,\,l}=\frac{Q^{2}\lambda_{m}^{2}\,l^{2}}{2m\left(2\left|l\right|+1\right)}.
\label{3.15}
\end{eqnarray}

In this way, the general expression for the energy levels (\ref{3.14}) is given by:
\begin{eqnarray}
\mathcal{E}_{n,\,l}=\omega_{n,\,l}\,\left[n+\left|l\right|+1\right]+\frac{k^{2}}{2m}.
\label{3.16}
\end{eqnarray}

Hence, we have seen in Eq. (\ref{3.16}) that the effects of the Coulomb-like potential induced by the interaction between the magnetic field and the electric quadrupole moment on the spectrum of energy of the harmonic oscillator corresponds to a change of the energy levels, where the ground state is defined by the quantum number $n=1$ and the angular frequency depends on the quantum numbers $\left\{n,\,l\right\}$. This dependence of the cyclotron frequency on the quantum numbers $\left\{n,l\right\}$ means that not all values of the cyclotron frequency are allowed, but a discrete set of values for the cyclotron frequency \cite{eug,f1,bb2,bm,bb4}.

\section{conclusions}

We have seen that bound states solutions for the Schr\"odinger equation arise from the interaction between a moving electric quadrupole moment and an external magnetic field. We have shown that the interaction between a moving electric quadrupole moment and a radial magnetic field can induce a Coulomb-like potential, where the spectrum of energy of the bound state solutions are defined for all values of the quantum number $l$ that differs from zero. For $l=0$, the term which plays the role of the attractive Coulomb-type potential vanishes and no bound states exist. 

Furthermore, we have studied the influence of the Coulomb-like potential induced by the interaction between a moving electric quadrupole moment and a radial magnetic field on the harmonic oscillator potential and shown that the ground state of the harmonic oscillator is defined by the quantum number $n=1$ instead of the well-known quantum number $n=0$. Moreover, we have shown that the spectrum of energy of the harmonic oscillator is defined for all values of the quantum number $l$ that differ from zero, that is, for $l\neq0$. For $l=0$, there is no influence of the Coulomb-like potential on the harmonic oscillator. Other effect of the Coulomb-like potential induced by the interaction between a moving electric quadrupole moment and a radial magnetic field on the harmonic oscillator is the dependence of the angular frequency on the quantum numbers $\left\{n,\,l\right\}$, which means that not all values of the cyclotron frequency are allowed. As example, we have calculated the angular frequency of the ground states $n=1$.

\acknowledgments

The author would like to thank CNPq (Conselho Nacional de Desenvolvimento Cient\'ifico e Tecnol\'ogico - Brazil) for financial support.

\end{document}